\documentclass[conference]{IEEEtran}
\ifCLASSOPTIONcompsoc
  \usepackage[nocompress]{cite}
\else
  \usepackage{cite}
\fi
\ifCLASSINFOpdf
\else
\fi
\usepackage{color,soul}
\usepackage{framed}
\usepackage{url}
\usepackage{graphicx}
\usepackage{multirow}
\usepackage[table,xcdraw]{xcolor}
\usepackage{fontawesome}

\newcommand\consorzio{Reclamation Consortium}

\newcommand\consorzioacronym{RC}

\begin{document}
%
\title{Eliciting the Double-edged Impact \\ of Digitalisation: a Case Study in Rural Areas}



%

\author{\IEEEauthorblockN{Alessio Ferrari\IEEEauthorrefmark{1},
Fabio Lepore\IEEEauthorrefmark{2},
Livia Ortolani\IEEEauthorrefmark{2} and 
Gianluca Brunori\IEEEauthorrefmark{2}}
\IEEEauthorblockA{\IEEEauthorrefmark{1} CNR-ISTI, 
Pisa, Italy\\ Email: alessio.ferrari@isti.cnr.it}
\IEEEauthorblockA{\IEEEauthorrefmark{2} University of Pisa, Department of Agriculture, Food and Environment,  
Pisa, Italy\\ Email: fabio.lepore@phd.unipi.it, livia.ortolani@agr.unipi.it, gianluca.brunori@unipi.it}
}


\maketitle


\begin{abstract}
Designing systems that account for sustainability concerns demands for a better understanding of the \textit{impact} that digital technology interventions can have on a certain socio-technical context. However, limited studies are available about the elicitation of impact-related information from stakeholders, and strategies are particularly needed to elicit possible long-term effects, including \textit{negative} ones, that go beyond the planned system goals. 
This paper reports a case study about the impact of digitalisation in remote mountain areas, in the context of a system for ordinary land management and hydro-geological risk control. The elicitation process was based on interviews and workshops. In the initial phase, past and present impacts were identified. In a second phase, future impacts were forecasted through the discussion of two alternative scenarios: a dystopic, technology-intensive one, and a technology-balanced one. The approach was particularly effective in identifying negative impacts.  
Among them, we highlight the higher stress due to the excess of connectivity, the partial reduction of decision-making abilities, and the risk of marginalisation for certain types of stakeholders. The study posits that before the elicitation of system goals, requirements engineers need to identify the socio-economic impacts of ICT technologies included in the system, as negative effects need to be properly mitigated. Our study contributes to the literature with: a set of impacts specific to the case, which can apply to similar contexts; an effective approach for impact elicitation; and a list of lessons learned from the experience. 
\end{abstract}

\IEEEpeerreviewmaketitle

\section{Introduction}
Sustainable innovation asks software engineers to reflect on the potential short- and long-term \textit{impacts} of their technical solutions in relation to the environment in which they are deployed~\cite{becker2015sustainability}. Requirements engineering (RE) is recognised as the key area to address sustainability concerns, and RE processes shall account for the identification of potential impacts of 
system development 
from the social, ecologic, and economic viewpoints~\cite{easterbrook2014computational,becker2015requirements}. This means going beyond the elicitation of system goals, which are the typical focus of RE, and embracing a wider perspective to identify the effects that a system can have on the environment in which it will be deployed. 

Several works have been carried out at the boundary of RE and sustainability~\cite{seyff2018tailoring,SAPUTRI2021106407,calero2015green,duboc2020requirements,garcia2018interactions,cabot2009integrating,phamBHM21,ferrario2016values,mahaux2011discovering,KERN2018199,doerr2018reinrural}. Part of them focus on energy-management  aspects~\cite{calero2015green,ferrario2016values,mahaux2011discovering,KERN2018199}, while 
others go beyond this perspective, and also consider the socio-economic dimension of sustainability~\cite{roher2013sustainability,seyff2018tailoring,SAPUTRI2021106407,lami2012measuring,bozzelli2013systematic,phamBHM21}. 



However, few studies have reported about the elicitation of the impacts of digital technology interventions in a real-world context. Furthermore, existing studies have encountered difficulties in eliciting \textit{negative} impacts~\cite{seyff2018tailoring,ferrari2022,duboc2020requirements}. Understanding the past, present, and envisioned impacts can provide useful information to prevent undesired consequences with future ICT solutions. 

To address this gap, this paper presents a case study about the impact of digital technology interventions on a socio-technical system in rural areas. Specifically, we study a system that supports ordinary land management and hydro-geological risk control in a scarcely populated mountain region. 
The system is currently composed of multiple private and public stakeholders and relies on the contribution of the inhabitants of the area. Designated inhabitants---forestry farmers and other subjects---can notify to a central authority the need for specific interventions, in relation, e.g., to floods or landslides. Then, the central authority (called Reclamation Consortium, RC) can assign the maintenance work to their own technicians, or to the inhabitants themselves. 
This system is going through a transformation in which the process is increasingly supported by digital technologies for communication, monitoring and data analysis.  
Our goal is to elicit the impacts that digitalisation has had in recent years, as well as envisioned ones.

To this end, we first performed a set of ten interviews and a workshop to elicit the \textit{experienced} impacts, followed by two additional workshops. The two additional workshops were oriented to brainstorming about the impact of digital technologies considering  \textit{technology-balanced} (desirable) and \textit{technology-intensive} (dystopic) scenarios. These polarised, yet realistic, scenarios were expected to better trigger contrasting visions. Overall, 35 subjects were involved in the study.

The elicitation process led to the identification of 16 positive and 18 negative classes of impacts. Informed by this knowledge, the socio-technical system can evolve taking into account the issues currently identified, thus minimising negative socio-economic impacts and maximising positive ones. Our study contributes with the practical experience of applying a RE process oriented to elicit impacts, and with a set of issues to consider when dealing with similar digitalisation contexts. 
Furthermore, we present take-away messages to support requirements engineers operating in a social environment. 

Raw data from the study cannot be shared for confidentiality reasons. However, we report an extensive set of quotes (translated from Italian), to ensure credibility and transparency.

The rest of the paper is structured as follows. Sect.~\ref{sec:background} describes the context of the study. Sect.~\ref{sec:design} describes the research design, and Sect.~\ref{sec:results} reports the results. Sect.~\ref{sec:discussion} discusses the main take-away messages and lessons learned. Sect.~\ref{sec:related} summarises related work and our contribution. Sect.~\ref{sec:limitations} presents limitations, and Sect.~\ref{sec:conclusion} concludes the paper.

\section{Context of the Study}
\label{sec:background}
In this section we provide a thick characterisation of the study context, as recommended by case study guidelines~\cite{runeson2012case}.

\paragraph{Socio-economic Context} The rural area considered in this study is located in northern Tuscany, Italy. It is characterized by a mountain landscape, it has an extension of 1,062.40 km\textsuperscript{2} and a total population of 37.312 inhabitants. Four out of the 19 municipalities included in the area are considered as ``ultra-marginal'' (over 75 minutes by car from larger towns) and the others 15 as ``marginal'' (over 40 minutes), according to the National Strategy for Inner Areas~\cite{innerarea2022}. Agriculture is one of the key economic activity, together with tourism. The main agricultural production in the area is vineyards and other tree production typical of mountain areas, together with honey production. Family gardens for vegetable production are also common land use in the area. Many farms have woods and forests attached to their production activities, and some of the inhabitants are forestry farmers, growing chestnuts, or raising native breeds of pigs and goats in the semi-wild state. The farmers of the area historically ensure the preservation of the ecological and hydro-geological stability of the area, through the maintenance of the riverbeds nearby their farms. 

\paragraph{Hydro-geological Risk} 
The maintenance of hydro-geological equilibrium is a key issue in Italy and in the specific area. 
The considered area is classified for the 70\% with a high and very high hydro-geological risk, and the level of seismic danger is one of the higher at the national level and the highest in the Tuscany region\cite{areeinterne2020}. 
The hydrogeological instability, due to the presence of waterbeds both on the ground and underground, is further exacerbated by erosion phenomena and landslides, which are frequent in the area. 


\paragraph{Depopulation} The structural fragility of this area, characterized by distance from essential services, has determined a long-term process of depopulation, aging of the population, and loss of key functions. 
In the last 15 years, the elder population increased by 2,5\% and the young population (0-14 years old) decreased by 12\%~\cite{areeinterne2020}. 
The depopulation has led to increased hydro-geological risk for the area, as the historic role of farmers in preserving the stability of the land through the constant maintenance works on the hydrographic system is being fulfilled by fewer and fewer people.  


\paragraph{Governance}
The geographical area has a local governance structure based on the coordination of some public services at the inter-municipal level.  
The Municipalities Unions are today the authority that provides the associate functions for tourism, civil protection, public works, and forest management. 
The specific competence for water resources management and safeguarding of water and hydro-geological risks belongs 
to the \consorzio{} (\consorzioacronym{}), 
which today performs this function both for mountain and plain areas.

\paragraph{Current Process} The \consorzioacronym{} recognises the professional experience of local farmers and their role in land management and reduction of hydrogeological risk. In particular, they aim to foster an active role of farmers in the alert system, as they live on the land and have a continuous observation perspective. Starting in 2006, a collaborative network was created between public institutions and around 30 farmers located in marginal mountain areas~\cite{vanni2014project}. These farmers, named ``custodian farmers'', receive incentives to monitor certain pre-defined areas, and notify possible criticalities (e.g., landslides, fallen trees) to the public authorities. Furthermore, through a public process, farmers can also participate in the solution of the identified criticalities, thanks to their equipment and their knowledge of the land. 
The farmers signal problems to the \consorzioacronym{} via WhatsApp and e-mails, and are assigned works through public calls for tenders. The \consorzioacronym{} wants to leverage these experiences to empower farmers, but also include citizens and tourists alike in the monitoring and management of the territory. The goal of the \consorzioacronym{} is to understand how this can be achieved through a systematic usage of digital technologies, e.g., drones, sensors, 5G connectivity, and artificial intelligence components for predictive maintenance. To support the \consorzioacronym{} in this endeavour, the research reported in this paper carries out a set of elicitation activities oriented to identify the current and future impacts of digital technologies. This output will be used in an activity of co-design of a specific technological solution for the area---activity not reported in the paper.

\section{Study Design}
\label{sec:design}


This study follows the case study approach according to Runeson \textit{et al.}~\cite{runeson2012case}, and considering the essential attributes required by the Empirical Standards~\cite{ralph2020empirical}. This type of research strategy is deemed appropriate to study a phenomenon within its real-life context, in which the boundary between the phenomenon and the environment cannot be clearly identified. 
The unit of analysis of the case study is the system for ordinary land management and hydrogeological risk control described in the previous section. The choice of this unit is opportunistic, as the authors are involved in a collaboration project with the \consorzioacronym{}. 
The following research questions (RQs) are addressed:



\textbf{RQ1:} \textit{What is the experienced impact of digital technologies from the viewpoint of the stakeholders of the socio-technical system?} With this question, we aim to elicit information about the perception of digital technologies, and how they have changed the work and daily life of the involved actors, from the individual, and socio-economic viewpoints. This can be particularly useful for understanding what are the potential issues related to digital technologies within the system-as-is. 

\textbf{RQ2:} \textit{What is the foreseen impact of digital technologies from the viewpoint of the stakeholders of the socio-technical system?} This question aims to identify possible expectations and fears with respect to changes in the system, triggered by the introduction of novel technologies. This can be useful for understanding undesired effects of digital technologies within the system-to-be.

To collect relevant data for RQ1, ten interviews and a workshop (\textit{impact workshop}) were conducted with representative stakeholders of the socio-technical system. The goal of the interviews was to have individual viewpoints on the experienced impact of digital technologies. The goal of the workshop was to compare viewpoints on the impacts, possibly identifying diverse opinions. To answer RQ2, two  workshops (\textit{scenario workshops}) were conducted in parallel. The first one (\textit{technology-balanced}) discussed possible impacts in a fictional scenario in which advanced technologies are widely used and there is a fruitful integration between social actors and ICT components. The second one (\textit{technology-intensive}) discussed possible impacts in case digital technologies are massively used, with limited human contribution.   



\subsection{Data Collection} 
 
\paragraph{Interviews}  
The interviewers are the first three authors. They have different backgrounds, namely software engineering (RE\#1), agricultural sciences (RE\#2), and social sciences (RE\#3). This heterogeneous group of interviewers was formed based on the rationale that the different facets of a socio-technical system require diverse expertise to be investigated. The interviewees were opportunistically selected based on a list of contacts provided by the coordinator of the RC. These are two ICT technicians of the RC, one ICT service provider, a political subject, two farmers, two field technicians, and the coordinator of the RC.
Interviews were unstructured, given the limited knowledge that we had about the individual contexts of the interviewees. Each interview had the objective of gathering information about: (1) the type of work performed by the subject; (2) the role  of the subject in the socio-technical system; (3) the past and present impact that the introduction of the digital technologies had on the system. 

The interviews were led by RE\#3, given her  experience in this type of task. The other interviewers could ask questions or clarifications, to cover other aspects of interests not touched by the lead interviewer. Interviews lasted about one hour on average, and were tape recorded. The RC coordinator was interviewed twice, so ten interviews were carried out. 

\paragraph{Impact Workshop} After the interviews, a workshop with additional participants was organised. The goal of the workshop was to collectively discuss the impact of digital technologies, to complement the information elicited in the first phase. The workshop was directed mainly to farmers, technicians, associations and local administrators involved in the ordinary management of the mountain territory. The meeting was also open to all citizens interested in contributing to the discussion. A flyer was redacted, and sent to potentially interested participants by the RC coordinator. 



The workshop was conducted online via Zoom.  
In total, 22 subjects with ICT, political/administrative and farmer profile, participated in the workshops. Specifically, 12 farmers, 3 managers and members of cooperatives, 2 service providers, 2 technicians of the RC, 1 technician of associations, 1 from the administration and the RC coordinator. After an initial presentation of the context by the RC coordinator, and by the organisers, a first interactive session was set-up, moderated by RE\#1, given his experience in workshops and focus groups. The moderator asked the participants to individually think about two questions:   \textit{What is the main technological change you have experienced in ordinary land management over the past 10 years?\footnote{This time span was agreed with the RC coordinator, considering the period when the most important technological changes started in the specific context.}}; \textit{How has this technological change affected your work?} 
He gave 5 minutes to reflect on their answers, and asked them to write them down. Then, he started a round table, asking each participant to present themselves to the others, read their answers, and freely comment on them. This way, everyone could express their opinion with limited bias from the others. Also, this simple exercise facilitated ice-breaking, especially for those who were less keen to express their viewpoint due to personal inclinations~\cite{de2018requirements}. The other participants were allowed to complement, support the vision, or object. 
 At the end of the session, the participants could freely discuss aspects that were not addressed, or that came to their mind later in the discussion. 
 This interactive session lasted about 1 hour and 30 minutes. 

After a break, a shorter interactive session was set-up, aimed at brainstorming about needs and expectations around digitalization. 
The session lasted about 30 minutes. 

\paragraph{Scenario Workshops} The scenario workshops involved a total of 22 people, belonging to different profiles---9 farmers, 2 common citizens, 4 administrative subjects, 4 technicians, 1 ICT provider, and 2 academics with expertise in rural areas. Part of the participants (13) was also involved in the previous workshop and interviews. The workshops were carried out as in-person  meetings. The participants were split into two equal groups, to carry out the technology-balanced scenario workshop moderated by RE\#3, and the technology-intensive scenario workshop, moderated by RE\#1. 
In these workshops, the moderator described a possible future scenario in a time-span of 10 years, and asked the participants to comment on opportunities and risks of the two scenarios. 
The two scenarios were described by the participants as shown in Table~\ref{tab:scenario}. The mentioned technologies were identified in agreement with the RC coordinator.  
A whiteboard was used to keep track of the observations, and provide an overview of the discussion to the participants. Each workshop lasted 60 minutes, and was followed by a plenary, wrap-up discussion of 30 minutes.

\begin{table*}[]
\centering
\resizebox{\textwidth}{!}{
\begin{tabular}{|l|p{7cm}|p{7cm}|}

\hline
\textbf{Dimension}                 & \textbf{Technology-balanced Scenario   (desirable)}                                                                                                                               & \textbf{Technology-intensive Scenario   (dystopic)}                                                                                                                                          \\ \hline \hline
\textit{Density of Population}     & The rural population is stable                                                                                                                                                    & The rural population decreases   significantly                                                                                                                                               \\ \hline
\textit{Level of Competence}       & The younger generations have a   level of computer skills that allow them to use the new technologies   available on their own                                                    & The remaining farmers have a   level of computer skills that does not allow them to use on their own the new   technologies available                                                        \\ \hline
\textit{Investments in Technology} & Public/private investments   enable the partial development of digital platforms                                                                                                  & Public/private investments enable the development of efficient digital platforms                                                                                                             \\ \hline
\textit{Connectivity}              & 5G Connectivity and broadband infrastructures are fully available even in rural areas                                                                                                & 5G Connectivity and broadband infrastructures are fully availalbe even in rural areas                                                                                                           \\ \hline
\textit{Technology Level}          & Technologies already available today such as sensors, weather booths, drones, digital maps, and public administration platforms help humans to better make sense of the situation of the territory, but interoperability is limited & Interoperability between  sensors, weather booths, drones, digital maps, public administration platforms, together with AI technology for analysis and prediction allow us to have constant monitoring of hydrogeological risk                                      \\ \hline
\textit{Technology Cost}           & The cost of technology decreases   but it is still necessary to integrate human and ICT systems                                                                                   & The cost of technology decreases   and allows us to monitor multiple environmental aspects (so that what is now done  by people who communicate possible problems, can be done by ICT tools) \\ \hline
\textit{Environmental Awareness}   & The young population is   interested in being involved in land management                                                                                                         & The population has no time or  interest in being involved in land management                                                                                                                 \\ \hline
\textit{Climatic Events}           & Extreme climatic events are   increasing and so is the need to constantly monitor the territory                                                                                   & Extreme climatic events are increasing and so is the need to constantly monitor the territory                                                                                                \\ \hline
\textit{Digitalisation Policies}   & Government resources, with the support of digitalisation policies, partially finance the development and   purchase of digital technologies for land management                   & Government resources, with  the support of digitalisation policies, fully finance the development and   purchase of digital technologies for land management                                 \\ \hline
\end{tabular}
}
\caption{Description of the two scenarios, desirable and dystopic.}
\label{tab:scenario}
\end{table*}



\subsection{Data Analysis}
The recordings of interviews and workshop were manually transcribed by RE\#2.  
Then, RE\#2 performed thematic analysis for RQ1 and RQ2 with open coding, and following an analytic style, according to the guidelines of Saldana~\cite{saldana2021coding}. Themes were discussed by all the authors in multiple iterations, were classified as positive or negative impacts, and were finally grouped into higher-level categories. For example, the fragment \textit{The first [technology used] was certainly the mobile phone, then came the photos and then the maps, which also allow us in our work to have immediate contact with the Consortium}, was coded with the theme ``quality of communication'', classified as a positive impact. The theme was then included in the category ``organisational'', together with other themes such as ``increased efficiency'' and ``transparency'', emerging from other fragments. In the following, we present the results of this analysis. 

\begin{figure*}[t]
\includegraphics[width=\textwidth]{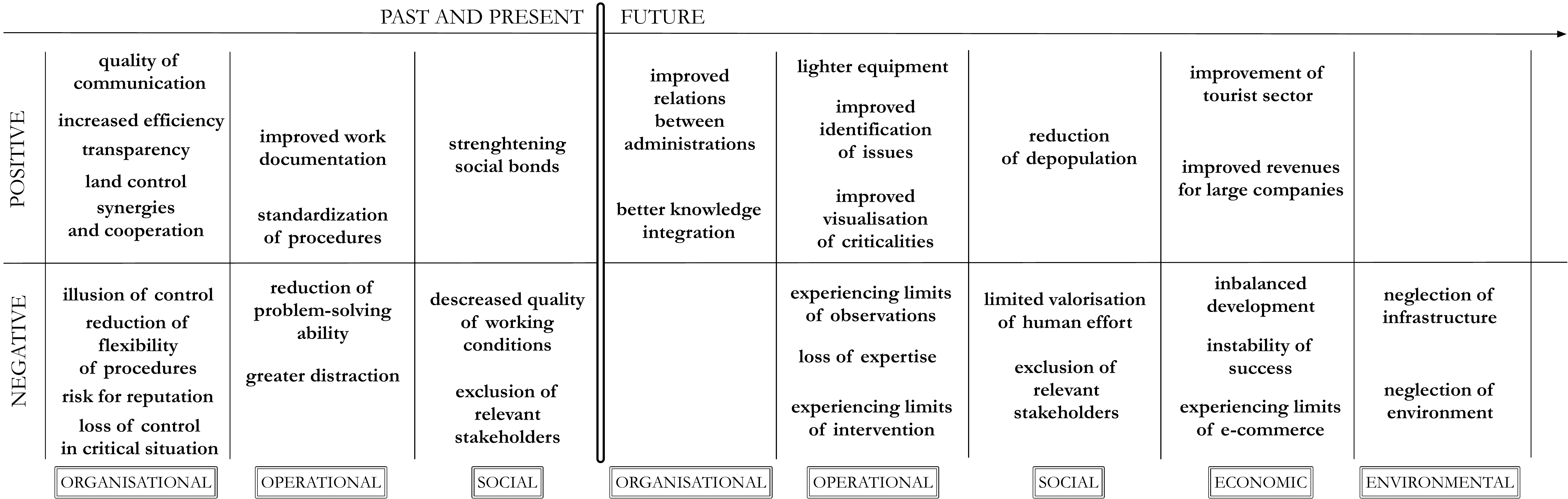}
\centering
\caption{Different positive and negative impacts, considering past and present, as well as future ones.}
\label{fig:impacts}
\end{figure*}

\section{Results}
\label{sec:results}

Fig.~\ref{fig:impacts} gives an overview of the different impacts, grouped by category. The categories are as follows: \textit{organisational} collects themes related to management and administration; \textit{operational} includes themes related to the technical interventions; \textit{social} includes relations, needs, and conditions of the involved stakeholders;  \textit{economic} collects economic- and revenue-related aspects; \textit{environmental} collect aspects related to infrastructures and natural resources. From the figure, it emerges a rather balanced view between positive and negative impacts, confirming that digitalisation is regarded as a double-edged phenomenon. We also see that the majority of the themes are categorised as operational and organisational, indicating that the focus of the stakeholders is on practical aspects, with less relevance given to social and environmental issues. This may be a limitation of our approach: taking a context-dependent, case study-based strategy may not be sufficient to elicit broader, higher-order impacts. Furthermore, while for the present and past impacts, economic and environmental aspects were not discussed, they became relevant when thinking about future perspectives. These appear pessimistic, given the majority of negative impacts for these two categories. In the following, we describe each impact for the different RQs. We list the classes of positive (+) and negative (-) impacts, together with representative quotes, also referring to the type of profile---Farmer, ICT, Political/Administrative (PA), or Field Technician (Field)---that stated the specific quote. 

\subsection{RQ1: What is the experienced impact of digital technologies from  the  viewpoint  of  the  stakeholders  of  the  socio-technical system?}



\textbf{+ Increased Efficiency.} Digital tools are recognised to improve the efficiency of the ordinary land management system. Digital maps and sensors allow technicians to have a larger availability of relevant data (hydrologic and climatic), reducing the need of on-site inspection, and enabling a better identification of the key areas that require maintenance works. 

\begin{small}
\textit{I find that the use of technology has improved our work organisation, reducing the need to travel for inspections or to deliver documents.} (Farmer)
\end{small}


\begin{small}
\textit{Simply using Google Maps allows me to arrive at a water course, locate it, see the access roads, and so I can also be quicker in achieving my goals and also organise my own and the other guys' working day.} (Farmer)
\end{small}





\textbf{+ Quality of Communication.} Communication quality improves in terms of speed and in terms of clarity, especially thanks to the multi-modality offered by current technologies. The usage of photos, videos, and the possibility of geo-referencing facilitates notifications. 

\begin{small}
\textit{What I have to say in my work in a general sense [...] is that the revolution is WhatsApp (or similar systems). That is, the possibility of exchanging with anyone any image, document, and even accompany them by messages, and speed.} (PA)
\end{small}

The possibility of using voice messages is also regarded as a major enabler of communication clarity.

\begin{small}
\textit{I spend two hours in the car in the morning and two hours in the evening, if I didn't use voice messages I don't think my performance in giving directions to colleagues would be any better. 
} (PA)
\end{small}

\textbf{+ Improved Work Documentation.} Using digital technologies also improves the documentation of the work. This is recognised especially by farmers who use pictures to document their work before, during and after interventions. The photos are used as visual notes to keep track of work and decisions. 

\begin{small}
\textit{With my mobile phone I use the classic WhatsApp to send photos before, during and after the intervention.} (Farmer)
\end{small}

\begin{small}
\textit{If they remain in your memory you remember what was said, what was done [...] So even when one refers to numbers and things, this information remains with you. I keep the messages just as visual notes and as an archive.} (Farmer)
\end{small}

\textbf{+ Land Control.} Greater efficiency is also associated with a feeling of having a better control of the territory. As some of the areas are scarcely populated, the communication of potential issues (e.g., need for removal of vegetation from waterbeds, presence of landslides in the street) by the few people who passes through them allows preventing hydro-geological risks. 

\begin{small}
\textit{With the photo of the area plus the geo-location of the street, citizens sent the report, and a good service was established 
to respond to the needs [...] So this is a change that has certainly brought benefits in terms of controlling the territory.} (PA)
\end{small}


\textbf{+ Synergies and Cooperation.} The control of the hydro-geological risk depends on multiple social and political actors, and responsibilities are not always clearly defined in public administrations. Technology-mediated communication has made it easier to communicate between public entities, and overcome problems of unclear regulations.

\begin{small}

\textit{There are several competent bodies and we don't make any effort to send an e-mail to colleagues in the municipalities [...]. 
Recently, we tried to take that extra step by forwarding the notifications [of issues]. Therefore, in my opinion, technology in shortening the distances between authorities [...], and it is an opportunity to fill those gaps that the law, regulations, and bureaucracy may have produced.} (PA)
\end{small}

\textbf{+ Standardization of Procedures.} Digital technologies are used to support the process of assigning maintenance works, and managing contracts at the administrative level. This is recognised to favour the homogenisation and standardisation of procedures between different offices of the territory. Furthermore, this also opens to the possibility of making statistics.

\begin{small}
\textit{
Everything that comes out in response from the Consortium must have standard patterns of response (it's not that Viareggio responds in one way and Capannori in another). [...] The other important thing is that in this way we can pull out statistics.} (ICT)
\end{small}

\textbf{+ Strengthening Social Bonds.} One indirect effect of technology adoption is the possibility of strengthening the bonds between family members, especially because many farm businesses are family-managed. Technology improves the links between generations, as clearly stated by the RC representative:

\begin{small}
\textit{Technology allows grandchildren to teach their grandparents and in my opinion this has a huge social value that should not be underestimated.} (PA)
\end{small}

\textbf{+ Transparency.} In previous years, a system was set-up, called \textsc{Idramap}, which allowed citizens to visualise the interventions made in the territory, thus following the evolution of the work, and having a feeling that their notifications were taken care of. This increased transparency reinforced the bond between the citizens and the RC.   

\begin{small}
\textit{But then there is also the issue of transparency, with citizens who always ask ``I pay the fee to the Reclamation Consortium, but what do they do with this money?''.  The Consortium's website started on the first page with a map showing all the work that was going on at the time} (ICT)
\end{small}

\textbf{- Reduced Problem-solving Ability.} The ease of communication also comes with the ease of delegation of decisions. This is considered to reduce the problem-solving ability of the workers. Decision are taken by people which are not on-site, thus reducing the quality of the solutions proposed.

\begin{small}
\textit{But it [technology] has inhibited the capacity for reasoning because whereas before if you had a group of people working and they found the difficulty, they would discuss it with each other, analyse the problem and make the decision, today this is no longer done because you find a difficulty, you take a picture and you wait for someone else to find the solution or tell you what to do.} (Farmer)
\end{small}

\textbf{- Illusion of control.} Although all the stakeholders celebrate the possibility of sharing and geo-tagging pictures, they agree on the fact that a photo is not enough to evaluate a situation, and can also give an illusion of control. 

\begin{small}
\textit{Certain things can only be assessed on site, 
you don't assess and quantify a job economically from a photo, because you can't understand it unless you were there the day before and so you know it perfectly. So I think there's an illusion that you can control everything from a distance, but that's not the case.} (PA) 
\end{small}

\textbf{- Loss of control in critical situations.} Excessive reliance on technology also leads to the risk of losing certain organisational functions when the technology does not work, which often happen in critical situations, as, e.g., snow storms.

\begin{small}
\textit{If we consider, for example, the recent snowfall in our area, all the [...] repeaters were down. So there is a risk that the more technology relies on slightly more sophisticated online systems [...], the more the communication system goes into crisis at critical moments.} (PA)
\end{small}

\textbf{- Greater Distraction.} Technology can be particularly distracting, and this can have negative impacts especially for manual workers.  

\begin{small}
\textit{How many people do you see on a crane with an earpiece? And it's a very dangerous thing, because it's true that you have the earpiece and you have the phone to your ear, but if you're talking to a person you're not paying attention to what the clamp is doing.} (Farmer)
\end{small}

\textbf{- Decreased  Quality  of  Working  Conditions.} Though in general the quality of the work is considered to be improved thanks to the increased efficiency, there is some imbalance in terms of working conditions between those that stay in their office, and those that stay in the field. The former can quickly and more easily communicate, but the latter are required to respond promptly to the requests, and it is not always easy.

\begin{small}
\textit{People in offices are in front of a computer, they open all the boxes, all the windows and send us documents. If we are in the woods, in the barn, driving a tractor in a field, it is much more difficult for us. [...] So while on the one hand it saves time, on the other hand it takes up a lot of time.} (Farmer)
\end{small}

Furthermore, digital communication also implies that disconnection is often impossible and this has effects on the working conditions of every stakeholder.

\begin{small}
\textit{But of course there's also the downside, because with mobile phones and e-mail you can always be tracked. There are no Saturdays and Sundays, but that's definitely the price you pay. 
} (Farmer)
\end{small}

Finally, ICT experts in public administration offices often have to do all kinds of ICT-related activities, often going far beyond their formal mansions. This creates an overload of tasks, which can impact the quality of the digital services.

\begin{small}
\textit{For example in Viareggio there is an engineer who takes care of maintenance, but at the same time he is responsible for the server room and other things, but he can't even manage everything} (ICT) 
\end{small}

\textbf{- Exclusion of Relevant Stakeholders.} One important aspect is the risk of excluding from the system that part of the population that can be particularly important for notifying certain issues, but is not sufficiently skilled with digital technologies. 

\begin{small}
\textit{Sometimes, 
I find myself in difficulty because I may ask them [farmers] to send me a position or to frame the area with the phone, and on the other side I have a person who does not respond to these requests because maybe they are not so super-digital.} (ICT)
\end{small}

However, the main risk of exclusion is due to the lack of connectivity in several areas, an issue that is remarked by all the stakeholders.

\begin{small}
\textit{We had asked the Region of Tuscany to speed up the project with Open Fiber broadband that had been confirmed by the end of 2020. Today, however, perhaps because of the COVID is moved to the end of 2021. Now we can be patient for another year, but frankly it is a penalty, a huge gap.} (Farmer)
\end{small}

\begin{small}
\textit{We pay for three subscriptions (one with the former EUTELIA, one with the phone line...which gives practically nothing, and one with a local operator via antenna), and despite this the bandwidth often runs out, so we can't make calls with WhatsApp.} (Farmer)
\end{small}

\textbf{- Reduction of Flexibility of Procedures.} The standardisation of procedures leads to problems when different offices are involved, and the governance is distributed. People have to change their usual way of working, and need to abide to procedures that are less adaptable than before. 

\begin{small}
\textit{The specifications [to digitalise the procedure] are based a little on how Capannori worked (without taking into account how they worked in Viareggio or Massa). [...] We had a bit of a problem because someone on the other sites was working differently and therefore had to adapt, so they also complained.} (ICT)
\end{small}

\textbf{- Risk for Reputation.} The increased transparency achieved through a system like \textsc{Idramap}, in which ongoing works are shown to citizens, also comes with the risk of possible criticisms. If certain works are delayed, this is made visible, and some political entities could be reluctant to introduce these innovations. 

\begin{small}
\textit{The fear that emerged even when we presented the project to the Municipality of [X] and the Municipality of [Y], was that there was a hostility to include what was being done, because there was a fear of exposing oneself to criticism.} (ICT)
\end{small}

\subsection{What is the foreseen impact of digital technologies
from the viewpoint of the stakeholders of the socio-technical
system?}

\textbf{+ Improved Relations between Administrations.} Currently, there is confusion about the roles of different administrative bodies belonging to the system. Technology can help to clarify responsibilities since it facilitates communication and standardisation. 

\begin{small}
\textit{We talked about the importance of interoperability, because there is a problem of confusion of roles and overlapping competencies
between different policies and different institutions. 
So the idea is that technology can also be used to facilitate this type of
relationships.} (PA)
\end{small}

\textbf{+ Better Knowledge Integration} Future technology is expected to integrate both human knowledge and data, coming from sensors, satellites, or other data sources.

\begin{small}
    \textit{This could also be a supplement to satellite data... it means putting your knowledge into an argument that is otherwise only made on the basis of machines.} (ICT)
\end{small}

\textbf{+ Lighter Equipment} Digital technology is expected to provide lighter equipment, which is particularly useful for workers, and technical staff who need to perform interventions. 

\begin{small}
    \textit{And as far as evolution is concerned
10 years from now, I have to say that I come from a mountain area, and
when you're in the mountains the fewer things you have on you the better [...].
having a tablet or something like that becomes a bit more challenging} (Field)
\end{small}

\textbf{+ Improved Identification of Issues} In the future, technology is expected to enable experts to better locate possible issues, by having specialised maps that are updated real-time according to the current situation. 

\begin{small}
    \textit{a small step further would be to
locate with the application on the map to intervene (partly with
partly by experience and partly by looking at the phone) immediately in the
right location.
} (Field)
\end{small}

\textbf{+ Improved Visualisation of Criticalities} To prevent issues, technology is also expected to enable the mapping of critical points, so that maintenance can be made before undesired events occur.

\begin{small}
    \textit{It would be useful [...] to have a mapping of critical points, i.e. to know the terrain, where I have to go to see, or maybe go to see when it rains.
} (Field)
\end{small}

\textbf{+ Reduction of Depopulation} The support of technology, with proper integration with human knowledge, can in the long term incentivise farmers to remain in the territory. Technology thus helps to give value to the humans that can exploit it. 

\begin{small}
\textit{We encourage pioneer mountain farmers to stay. Then clearly they must be trained, but they must also be economically incentivised to stay. And the contribution that is made through labour, in a mountain farm budget, is incredibly valuable.} (PA)
\end{small}

\textbf{+ Improvement of Tourist Sector} Technology, including platforms for booking holidays, as well as the availability of internet connection, can increase the chance of attracting tourists, thus helping  operators in the tourism sector.  

\begin{small}
\textit{
I have two agritourisms, and in my opinion, technology for tourism is very important to make even marginal areas known.
} (Farmer)
\end{small}

\textbf{- Imbalanced Development.} Public funds are increasingly available for digitalisation. This creates imbalance between different people with different educational levels, especially regarding the access to funds.

\begin{small}
    \textit{The winners are those who have a high educational level, a high economic availability that allows them to know and exploit the opportunities that politics generates. If, on the other hand, you are not educated, collect chestnuts or are a woodcutter, you are a loser.} (Farmer)
\end{small}

\textbf{- Instability of Success.} Even the well-educated subjects who are able to exploit digitalisation funds pursue short-term objectives and do not contribute to long-term goals. Therefore their success and the overall system sustainability is not long-lasting.

\begin{small}
    \textit{I would also add that these good people here, who find the right channel, have an upstream myopia related to the fact that this thing is very short. The success of this is very short, it is not the development for the future.} (Farmer)
\end{small}

\textbf{- Experiencing Limits of E-commerce.} While e-commerce can help the diffusion of local products, it is not appropriate for fresh products. Furthermore, small business have limited amounts of products and cannot satisfy the wide market that e-commerce can create. 

\begin{small}
    \textit{[E-commerce] has given the opportunity to sell a local product to a wider market [...] but the quantity is really negligible as far as our products are concerned. [...] I would be afraid to set up an e-commerce, because once I'm in, the demands overwhelm me.} (Farmer)
\end{small}

\textbf{- Neglection of Infrastructure.} When focusing on ICT development, one should not forget to ensure the connectivity infrastructure, but also physical infrastructures such as streets, to let the system develop from the digital and physical standpoints. (Field)

\begin{small}
    \textit{I would have asked myself what needs to be done to ensure that in 10 years there is no one left. Infrastructure, roads, fibre optics must be worked on.}
\end{small}

\textbf{- Neglection of Environment.} When focusing on ICT and infrastructure development, one risk is to neglect the need to ensure maintenance of the natural environment, which is crucial to avoid floods in this context. 

\begin{small}
    \textit{One concrete and necessary thing in Garfagnana, Lunigiana and mountain areas in general is to clear the forests to keep them as they were kept by our elders. Because this is one of the causes of flooding.} (Field)
\end{small}

\textbf{- Limited Valorisation of Human Effort.} While most of the funds coming from administrations are oriented to facilitate digitalisation, there is the need to give the right value and appropriate funding to the people who live and work in the area, and are the custodians of the territory. 

\begin{small}
    \textit{It is those of us who work on the territory (and therefore our work is a service to the territory) who cannot do it alone, but need investment [...] which are useful to keep the population there and prevent damage. } (Farmer)
\end{small}

\textbf{- Experiencing Limitations of Observations.} While sensors and drones can be useful to observe and monitor the territory, they cannot be used in all contexts, especially in forests and all areas in which the visibility is limited. Excessive expectation of technology will lead to coming to terms with its limitations.

\begin{small}
   \textit{Flat and easily accessible areas you can fly over with a drone, you can put sensors and weather stations. But if you look at the Apuan Alps, the whole hilly and mountainous area, there is very little you can see with drones.} (Field)
\end{small}

\textbf{- Experiencing Limitations of Intervention.} While in the near future one can expect certain robots to perform maintenance on easily accessible areas, the preservation of more remote areas will still require human intervention.

\begin{small}
    \textit{Maybe in the accessible areas it might be that a robot goes (as does a flail mower), but up there how do you do it? The grass has to be cut by hand.} (Field)
\end{small}

\textbf{- Loss of Expertise.} When using automated solutions, there is the risk of losing the specific expertise needed to perform interventions, and understand the situation in specific areas of the territory. 

\begin{small}
    \textit{I am talking about the data of
knowledge of the territory, which may be lost,
because it is knowledge that maybe farmers know,
hunters know, people who normally go into the territory know. So
know that when it rains a lot you have to go and see that bridge
there, or that bend in the river.} (Field)
\end{small}

\section{Discussion and Lessons Learned}
\label{sec:discussion}
The presented experience led to a set of lessons learned, which can be useful for readers dealing with similar contexts. Specifically, our conclusions can be helpful to requirements engineers (and researchers) developing (or studying)  socio-technical information systems that: involve public actors, private companies, and citizens belonging to rural communities; strongly rely on technology-mediated communication between actors; need to consider regulations and procedures; include a cyber-physical dimension (e.g., sensors). 

In the following, we list our reflections, and link them to related literature in RE. For each reflection, we also highlight the main implications with the \faLightbulbO{} symbol.

\paragraph{The Importance of Eliciting Impacts in RE} Requirements elicitation activities are commonly focused on the elicitation of \textit{goals}, i.e., properties of the environment that one wishes to satisfy (or satisfice) with the development of the system~\cite{van2009requirements,zave1997four}. With the introduction of goal-oriented methods to model socially-intensive systems, greater relevance has been given to \textit{stakeholders'} goals, as, e.g., in i*~\cite{eric2011social}. However, the focus of requirements elicitation remained the identification of objectives that actors may \textit{voluntarily} want to pursue with the development of a certain system. In this paper, together with other authors~\cite{becker2015sustainability,ferrari2022}, we posit that, besides system or stakeholders' goals, socio-technical system development and evolution through digitalisation needs to consider technological \textit{impacts} beforehand. 
The concept of impact is analogous to that already considered, among others, by Brito \textit{et al.}~\cite{brito2018} and by Seyff \textit{et al.}~\cite{seyff2018tailoring}, and is intended as the expected effect that a digital technology can have from a sustainability standpoint, and thus in mid- to long-term. While goals are typically positive, impacts can also be negative, and need to be thoroughly investigated in advance during the early RE phases, especially when socio-economic, organisational, and operational dimensions are involved. 
Professional requirements engineers should take into account our identified impacts during requirements definition,
so that undesired effects can be mitigated. In particular, we highlight the following important recommendations, derived from our themes:

\faLightbulbO{} Complementary, low-technology procedures should be introduced to cope with possible unreliability of technology (cf. theme \textit{loss of control in critical situations}). 

\faLightbulbO{} Technology-supported procedures should be adaptable to different contextual variants to avoid  strict constraints (\textit{reduction of flexibility of procedures}).

\faLightbulbO{} Enforcing process transparency towards users shall also consider the risk of compromising reputation of the information producers. Thus, transparency shall be always coupled with clear explanations/justification of the exhibited behaviour (\textit{risk for reputation}).

\faLightbulbO{} System development at the administrative level shall be coupled with appropriate policies and solutions that ensure the inclusion of all relevant stakeholders, as these can provide important information for the system (\textit{exclusion of relevant stakeholders}).

\faLightbulbO{} From the operational standpoint, one shall prefer solutions in which humans and ICT systems cooperate, instead of technology-intensive ones, as certain operations can be performed only by domain experts. Besides the inherent limits of technology, the tacit knowledge of experts is hard to encode in ICT solutions~\cite{ferrariSG16,gervasi2013unpacking}.  (\textit{experiencing limits of observations/ intervention}, \textit{loss of expertise}). 

\faLightbulbO{} When integrating ICT solutions in human-intensive systems, especially safety-related ones, one needs to consider the information overload and distraction triggered by the ICT system. Therefore, procedure shall be established to minimise human-machine interaction. (\textit{greater distraction} \textit{reduced problem-solving ability}).

\faLightbulbO{} From the economic standpoint, ICT solutions shall also account for small players, so that digitalisation does not favour large organisations only. Lightweight and tailored tools shall be considered that include all the possible subjects in the economic growth triggered by the system, so that socio-economic sustainability is ensured (\textit{imbalanced development, instability of success, experiencing limits of e-commerce}).

\faLightbulbO{} When focusing on ICT systems, requirements engineers shall not lose the focus on physical aspects, besides human ones. Physical infrastructure and environment play a crucial role in the achievement of system goals, and joint procedures shall be established so that the physical dimension is not neglected (\textit{neglection of infrastructure/environment}).

\paragraph{Eliciting Negative Impacts} Previous literature has stressed the difficulty of eliciting negative impacts from stakeholders~\cite{seyff2018tailoring,ferrari2022,duboc2020requirements}. 
With our analysis, which looks at past and present experience, as well as at future scenarios, we were able to list 18 general classes of negative impacts. Among them, only \textit{exclusion} was already captured by the study of Ferrari \textit{et al.}~\cite{ferrari2022}, who also focuses on rural areas and is based on 30 interviews with experts across EU. We also notice that the general taxonomy of Rolandi \textit{et al.}~\cite{su13095172} focuses on positive consequences. Therefore, we argue that looking at specific contexts, and asking stakeholders to reflect on their past experience, can be a simple, yet useful trigger to better identify negative impacts. Furthermore, an essential role was played by the discussion of the technology-intensive scenario. Indeed, the technology-balanced scenario mainly triggered reflections on how to overcome current barriers (e.g., improvement of communication infrastructure, novel administrative procedures) to achieve what looked as a desirable future in which the human role is not neglected. Instead, the technology-intensive scenario enabled people to think about a reality in which humans are not at the center; a somewhat dystopic future governed by machines. This scenario allowed us to elicit potential, and yet realistic, negative impacts that shall be prevented when introducing novel ICT-based systems in the specific socio-technical context. Researchers are thus called to further extend the list of impacts---and the scope of validity of the research---through other case studies. To this end, we recommend following the approach adopted in this paper: a set of interviews with relevant stakeholders, followed by one workshop to elicit reflections on present and past impacts, and by two separate ones exploring technology-balanced and -intensive scenarios. 

\faLightbulbO{} Identification of negative impacts can be performed by asking stakeholders to reflect on present and past, instead of solely brainstorming on \textit{future} effects.

\faLightbulbO{} When reflecting on future negative effects, it can be useful to depict a technology-intensive scenario in which humanity has a marginal role, and let participants brainstorm on what it would be to live in such a reality.

\faLightbulbO{}
The case study research strategy maximises the relevance of context~\cite{stol2018abc}, and it has the right reality-focused approach to identify the negative impacts of technology.


\paragraph{Liquid Nature of the System} One of the main difficulties that we encountered was understanding the roles of the participants and their relationships, as the organisation chart is liquid in these contexts in which public and private entities collaborate, both as individuals (e.g., farmers, single municipalities), and as groups (cooperatives of farmers, union of municipalities). This problem also occurs at the technical level, in which several databases and platforms co-exist, often not communicating with each other. It is also difficult to establish a boundary for the overall system, as some digital platforms (e.g., administrative databases) are used not solely in the context of ordinary land management (i.e, the main purpose of our system), but also for other aspects that are under the responsibility of the involved actors. To support understanding, and possibly modelling, uncertainty in  software and cyber-physical systems, RE techniques have been proposed~\cite{salay2013managing,zhang2018specifying,troya2021uncertainty}. We argue that such approaches need to be properly adapted for social-intensive systems as well, to ensure that uncertainty is properly captured. 

\faLightbulbO{}
RE researchers are called to refine domain modelling and requirements elicitation strategies, to take into account knowledge uncertainty in the study of social-intensive systems.   


\paragraph{Dynamic Nature of the System} When discussing about the current state of the socio-technical system and its impacts, it was not always clear which elements were currently part of the system, which elements did not belong to the system anymore, and which ones were going to be deployed in the near future. In this sense, we realised that we were capturing a fading picture of the system, due to its constant evolution over time. The evolution was at every level, including institutional, regulatory and technical ones. 
The need for RE practitioners to deal with a changing environment, and accept its inconsistencies has been largely stressed~\cite{ernst2014overview,ferrariSG16,nuseibeh2001making}. This attitude becomes crucial to address these contexts, whose inherently dynamic nature makes it impossible to actually engineer the system. This exists beforehand, its changes depend on multiple factors, and requirements engineers can only adjust or tune it. 

\faLightbulbO{} RE researchers need to define novel requirements analysis  strategies to allow capturing the changing nature of the systems under analysis, when these need to be re-engineered. 

\paragraph{Social Bonds and Trust as a Motivation} A primary role in indicating people to be interviewed, and contacting workshop participants was played by the RC coordinator. In a distributed social context, the relation of trust established across the years by the RC coordinator, the farmers and the other subjects made it possible to involve several people, without a direct reward. Someone even participated while staying outside in the snow with their mobile phone, as the indoor network coverage was not sufficient. It is therefore crucial to leverage existing trust relationships when studying these types of systems, as social bonds can act as incentives \textit{per se}. The classical RE problem of stakeholder identification~\cite{pacheco2012systematic,sharp1999stakeholder} is therefore intertwined with the design of public participatory processes~\cite{bryson2013designing}, in which participants need to be \textit{motivated}. This is in line with the observations of Kolpondinos and Glinz about the issues with reaching out stakeholders outside of organisations~\cite{kolpondinos2020garuso}, and ongoing discussions in Crowd RE~\cite{khan2019crowd,wouters2021crowdre}. In our specific context, we were not able to reach out big players, as e.g., large contractors, and high-rank political subjects. Different motivation strategies need to be devised to include these type of  stakeholders.  

\faLightbulbO{} When studying distributed  socio-technical systems, requirements engineers need to involve pivotal subjects with an established trust relationship with the other stakeholders. 

\faLightbulbO{} Novel stakeholder involvement strategies are required to attract big business players and political subjects, to successfully perform RE activities in socio-technical contexts.
\paragraph{The Importance of a Multi-disciplinary Team} Our group composition takes into account the recognised relevance of domain knowledge in requirements elicitation~\cite{hadarSK14,ferrariSG16}, as well as the need to include also domain ignorants~\cite{mehrotraB21} in the group, to facilitate elicitation of tacit knowledge and missing aspects~\cite{ferrariSG16,bano2019teaching}. Specifically, as interviewees included ICT, political subjects, and forestry farmers, individual interviewers had a different degree of expertise (and ignorance) in each interview/workshop. We recognise that this was crucial to capture the multiple facets of the system, especially given the different jargon used by the interviewees. For example, a participant with a political background used sentences like \textit{The National Strategy is made up of a document which is the strategy, plus a series of action sheets that activate the objectives listed in the Strategy. Each intervention sheet refers either to the funds of the National Strategy for Inner Areas (for that cohesion fund), or to the measures that are in the various PORs}. Instead, an ICT expert used expressions like \textit{REST API}, \textit{QFiled}, \textit{Debian 8}, \textit{VSDL}. Capturing information and merging these diverse data in a coherent view would hardly be possible with a team with a single background. 

\faLightbulbO{} Subjects with complementary and multi-disciplinary backgrounds need to be involved when impact elicitation needs to be carried out for complex socio-technical systems.

\section{Related Work}
\label{sec:related}

This paper belongs to the recent stream of literature about RE and sustainability~\cite{garcia2018interactions,venters2017characterising}. Within this line of research, we are mostly concerned with the \textit{socio-economic} side of sustainability. Indeed, it has been recognised that the introduction of digital technologies in socio-technical contexts can play a double-edged role, by privileging some stakeholders, and marginalising those who cannot cope with the change~\cite{herrero2021articulating,ruralstudies}.  Furthermore, ICT solutions may address short-term, stakeholder-specific goals, without considering potential impacts on individuals and society~\cite{ferrari2022,becker2015sustainability}.

To support RE practitioners in the development of systems that take sustainability aspects into account, Cabot \textit{et al.}~\cite{cabot2009integrating} propose to use the $i^*$  framework to make explicit the impact of each business and design alternative. Mussbacher \textit{et al.}~\cite{mussbacher2014goal} extends the Goal-oriented Requirements Language (GLR) to quantitatively evaluate the degree of sustainability of an envisioned system. Roher and Richardson~\cite{roher2013sustainability} proposes to reuse certain requirements patterns that are specifically concerned with sustainability aspects. Brito \textit{et al.}~\cite{brito2018} use aspect-oriented requirements analysis, and explicitly account for the potential sustainability-related effect of system requirements. Seyff \textit{et al.}~\cite{seyff2018tailoring} tailor the Win Win negotiation strategy to elicit sustainability-related impacts. By means of a case study, they show that it is particularly difficult for stakeholders to anticipate long-term effects. 
To better support this goal, Duboc \textit{et al.}~\cite{duboc2020requirements} present a set of questions to be asked to relevant actors, as well as a graphical notation, to facilitate reasoning about the potential short- and long-term impacts of an envisioned system. Saputri \textit{et al.}~\cite{saputri2020addressing} introduces a complete RE framework, which also includes specific metrics to evaluate sustainability aspects.  With a focus on rural areas, Ferrari \textit{et al.}~\cite{ferrari2022} presents a set of impacts of digital technologies in this domain, based on a set of interviews with experts. In the same field, Rolandi \textit{et al.}~\cite{su13095172} presents a taxonomy of impacts derived from a survey of the literature. Surveys concerning the interplay between RE and sustainability have been published considering scientific literature on the topic~\cite{garcia2018interactions}, and also based on interviews~\cite{chitchyan2016sustainability} and questionnaires~\cite{condori2018characterizing}. 

\textit{Contribution.} Our work presents a case study~\cite{runeson2012case} about the impact of digital technologies in a scarcely populated mountain region. With respect to previous works oriented to categorise impacts such as those by Ferrari \textit{et al.}~\cite{ferrari2022} and Rolandi \textit{et al.}~\cite{su13095172}, this work takes a context-dependent perspective, thus complementing the existing categorisation frameworks. Compared to works that use RE techniques~\cite{seyff2018tailoring,cabot2009integrating,mussbacher2014goal,brito2018,roher2013sustainability} or propose full-fledged frameworks~\cite{saputri2020addressing,duboc2020requirements} 
our study complements existing proposals with a simple, yet effective elicitation approach applied to a case study. 

We show that this different perspective allows us to identify concrete negative impacts, which are typically hard to elicit~\cite{seyff2018tailoring,ferrari2022,duboc2020requirements}. Our catalogue of impacts, and the approach used to identify them, 
can provide a baseline for future system analysis and development in similar contexts. 

\section{Threats to Validity}
\label{sec:limitations}
The case study is designed to fulfill the essential attributes required by the Empirical Standards~\cite{ralph2020empirical}. Limitations are discussed according to the categories illustrated by Leung~\cite{leung2015validity}.

\textit{Validity.} To guarantee the validity of our qualitative analysis, we used established practices of thematic analysis using the guidelines for coding by Salda{\~n}a~\cite{saldana2021coding}. Our study involved ~35 opportunistically selected subjects (part of the subjects participated in multiple activities), and we could have missed some relevant stakeholders. To mitigate this aspect, we reached out participants covering different roles. At the same time, some relevant roles could not be involved, and the representation is unavoidably partial, especially because \textit{saturation}~\cite{hoda2021socio} could not be  reached. 
However, we also argue that the overall goal of understanding the impact of digitalisation is reasonably addressed even with an incomplete set of stakeholders. We could not perform member-checking of the participants, to ensure that the information was correctly conveyed. However, our mix of backgrounds, and the use of tape-recording with manual transcription partially mitigate this issue.

\textit{Reliability.} Reliability of the analysis is ensured through triangulation between the different researchers. Specifically, the analysis of  RE\#2 was revised by RE\#1 and  RE\#3, in multiple meetings to come to a consolidated output. We do not compute, nor report the inter-rater reliability, as we did not perform an independent coding activity. As common for qualitative studies, the raw transcripts cannot be shared for confidentiality reasons. We however reported an extensive collection of excerpts of our themes (translated from Italian) to show the chain of evidence. Furthermore, we provided a thick characterisation of the study context to improve credibility, as recommended for case study research~\cite{runeson2012case,ralph2020empirical}.  

\textit{Generalisability.} As a case study, generalisability is inherently limited~\cite{stol2018abc}. According to case-based generalisation~\cite{wieringa2015six}, our results can be applied to similar contexts, where similarity is characterised by: types of actors (public, private, citizens); system type (distributed,  socio-technical, communication-intensive); domain (rural areas).  

\section{Conclusion}
\label{sec:conclusion}

This paper presents a qualitative case study about the impact of digital technologies interventions in remote mountain areas. 
We first elicit experienced impacts of digital tools currently used in the system, including instant messaging, and administrative platforms. In addition, we identify the possible impacts envisioned by the stakeholders when additional ICT technologies are introduced. 
In future work, we will use the collected impacts to inform the participatory design of a novel context-specific solution for hydrogeological risk control that takes into account the undesired long-term consequences of digitalisation that emerged in this study. 

As a personal remark of the authors, we were impressed by the high degree of competence and commitment of all the participants. One of the farmers said: \textit{My mind is now structured on digital. Every service nowadays (even for employees) goes through platforms. [...] Even my father, who was born in the 1950s, can do a lot with his mobile phone and apps.} Contrary to what has been observed by other authors~\cite{ferrari2022}, lack of skills does not emerge as a major issue in this context, as ``remote'' and ``rural'' do not imply digitally illiterate. It is the infrastructure that is lacking, not the skills, and policy makers and investors can play a crucial role in encouraging the expression of this potential. 

\bibliographystyle{IEEEtran}
\bibliography{IEEEabrv,bibliography.bib}
%

\end{document}